%% file: draft_A4-V4.tex
\documentclass[aps,prd,nofootinbib,floatfix,superscriptaddress,preprintnumbers,showpacs]{revtex4}
\usepackage{mathrsfs}
\usepackage{amsmath}
\usepackage{amssymb}
\usepackage{epsfig}
\usepackage{graphicx}
\usepackage{color}
\usepackage{subfigure} 
\include{bookmacros}

\DeclareMathAlphabet{\mathsc}{OT1}{cmr}{m}{sc}

\def\10{$SO(10)$}
\def\21{SU(2) $\otimes$ U(1) }

\def\422{$SU(4) \otimes SU(2) \otimes SU(2)$}
\def\321{SU(3) $\otimes$ SU(2) $\otimes$ U(1)}
\def\gsim{\raise0.3ex\hbox{$\;>$\kern-0.75em\raise-1.1ex\hbox{$\sim\;$}}}
\def\lsim{\raise0.3ex\hbox{$\;<$\kern-0.75em\raise-1.1ex\hbox{$\sim\;$}}}
\def\lsim{\raise0.3ex\hbox{$\;<$\kern-0.75em\raise-1.1ex\hbox{$\sim\;$}}}
\def\gsim{\raise0.3ex\hbox{$\;>$\kern-0.75em\raise-1.1ex\hbox{$\sim\;$}}}
\def\vev#1{\left\langle #1\right\rangle}
\def \znbb {0\nu\beta\beta}

\newcommand{\Addrfisteo}{%
  Departament de F\'{\i}sica Te\`{o}rica \& IFIC, 
  Universitat de Val\`encia/C.S.I.C.,\\
  Edificio Institutos de Paterna, Apt 22085, E--46071 Valencia, Spain}

\newcommand{\AddrFrascati}{%
  INFN, Laboratori Nazionali di Frascati,
Via Enrico Fermi 40, I-00044 Frascati, Italy}

\newcommand{\AddrWurzburg}{%
  Institut f{\"u}r Theoretische Physik und Astrophysik,\\
  Universit{\"a}t W{\"u}rzburg, 97074 W{\"u}rzburg, Germany}

\relax

\let\vev\VEV

\def \znbb {$0\nu\beta\beta$ }

\begin{document}

\title{\bf Quark-Lepton Mass Relation and CKM mixing in an $A_4$ Extension\\ of the Minimal Supersymmetric Standard
Model} 
 
\author{S. Morisi} \email{stefano.morisi@gmail.com} \affiliation{\AddrWurzburg}
\author{M. Nebot} \email{nebot@ific.uv.es} \affiliation{\Addrfisteo}
\author{Ketan M. Patel} \email{ketan@theory.tifr.res.in}
\affiliation{Department of Theoretical Physics, Tata Institute of Fundamental Research, Mumbai 400
005, India}  
\author{E. Peinado} \email{epeinado@ific.uv.es}\affiliation{\AddrFrascati} 
\author{J. W. F. Valle} \homepage{http://astroparticles.ific.uv.es/}
  \email{valle@ific.uv.es} 
\affiliation{\Addrfisteo} \date{\today}

\begin{abstract}
An interesting mass relation between down type quarks and charged leptons has been recently predicted within a
supersymmetric \SM model based on the $A_4$ flavor symmetry. Here we propose a simple extension which provides an
adequate full description of the quark sector. By adding a pair of vector-like up-quarks we show how the CKM entries
$V_{ub}$, $V_{cb}$, $V_{td}$ and $V_{ts}$ arise from deviations of the unitarity. We perform an analysis including
the most relevant observables in the quark sector, such as oscillations and rare decays of kaons, $B_d$ and $B_s$
mesons. In the lepton sector, model predicts an inverted hierarchy for the neutrino masses leading to a potentially
observable rate of neutrinoless double beta decay.
\end{abstract}

\preprint{IFIC/13-17}
\preprint{TIFR/TH/13-07}

\maketitle

\section{Introduction}

Understanding the observed pattern of quark and lepton masses and mixing from first principles constitutes one of the
deepest challenges in particle physics~\cite{Beringer:1900zz}. The recent robust experimental discovery of a nonzero
value for the reactor mixing angle $\theta_{13}$ in the neutrino sector~\cite{Tortola:2012te} may unveil surprises in
the underlying theoretical structure of the flavour sector~\cite{Morisi:2012fg}, opening also the door towards a new
generation of experiments searching for CP violation in the leptons~\cite{bandyopadhyay:2007kx,nunokawa:2007qh}. An 
ever growing body of experimental evidence makes flavour in the quark sector a challenging playground for any extension
of the Standard Model (SM).

Flavor symmetries provide a very useful approach towards reducing the
number of free parameters describing the structure of the fermion
sector. Non-Abelian discrete groups have played an important role in
connection with the flavour problem. Their mathematics is somewhat
less familiar to particle physicists as continuous non-Abelian
symmetries, and have been extensively discussed in the recent book,
Ref.~\cite{ishimori2012introduction}. Groups like
$A_4$\,\footnote{$A_4$ is the group of even permutation of four
  objects, for pioneering work
  see\,\cite{Babu:2002dz,Altarelli:2005yp}.} are especially useful in
that they contain triplet irreducible representations, exactly the
number of \SM generations.

In Ref.~\cite{Morisi:2011pt} a supersymmetric extension of the
standard \SM has been proposed based on the $A_4$ group, where all the
matter fields as well as the Higgs doublets were assigned to the same
$A_4$ representation, namely, the triplet. This leads to an important
theoretical prediction, namely a mass relation
\begin{equation}
\label{eq:ours}
\frac{m_{\tau}}{\sqrt{m_em_\mu}}\approx\frac{m_b}{\sqrt{m_d m_s}}~,
\end{equation}
involving down-type quarks and charged lepton mass ratios. Such relation provides a generalization of the three
Georgi-Jarlskog (GJ) mass relations~\cite{Georgi:1979df},
\begin{equation}\label{GJ}
\begin{array}{lll}
m_b=m_\tau,&m_s=m_\mu /3 ,&m_d=3 m_e,
\end{array}
\end{equation}
which arise within a particular ansatz for the SU(5) model and hold at the unification scale. In contrast to Eq.
(\ref{GJ}), our relation requires no unification group and holds at the electroweak scale. It would, in any case, be
rather robust against renormalization effects as it involves only mass ratios \footnote{The mass relation in Eq.
(\ref{eq:ours}) can get modified due to finite supersymmetric threshold corrections to the fermion masses. However such
corrections crucially depend on several details of the soft supersymmetry breaking parameters which are not constrained
by the model presented here. In order to keep our discussion as model-independent as possible we defer the discussion of
the details of soft supersymmetry breaking threshold effects to another publication.}.

A second prediction obtained in Ref.~\cite{Morisi:2011pt} involves the
Cabibbo angle for the quarks which arises mainly from the down-type
quark sector~\cite{Gatto:1968ss} with a correction coming from the up
isospin diagonalization matrix.  While this provides a successful
prediction for the Cabibbo angle the corresponding predictions for
$V_{ub}$, $V_{cb}$, $V_{td}$ and $V_{ts}$ were unacceptably small and
require an extension.

Following the suggestion in the original paper~\cite{Morisi:2011pt}
here we address such a possibility.  Recently some of us considered
the same problem in \cite{King:2013hj} where the CKM matrix is
obtained by assuming a different $A_4$ quark assignment. In this paper
we consider a variant of the first scheme in which extra fermions are
added, namely a pair of up-type quarks, in which case the full CKM
matrix will be a $4\times 3$ matrix. The small $V_{ij}$ mixings can be
generated by means of violations of $3\times 3$ unitarity. In the next
section we introduce our model, giving a brief description of quark
mixing as well as neutrino masses and mixing, in section
\ref{sec:phen-quark-sect} we address in detail the phenomenology of
the quark sector of the model, and in section~\ref{sec:conclusion} we
give our summary and conclusions.

\section{The Model}
\label{sec:model}

The basic content in terms of MSSM matter and Higgs superfields is the
same as that in \cite{Morisi:2011pt}. It is assumed that all such
fields transform as $A_4$ triplets. In order to generate correct
quark mixing angles without changing the mass relations between down
quarks and charged leptons, we introduce a pair of vector-like up-type
quark multiplets $T$ and $T^c$ transforming as $(3,1,2/3)$ and
$(\bar{3},1,-2/3)$ under the SM gauge group \SM.
In addition, we introduce two $A_4$ triplet flavon fields $\sigma$ and
$\sigma'$ imposing a $Z_4$ symmetry (with $\omega^4=1$) in order to
distinguish between them. The various charge assignments are given in
Table~\ref{tabf} where
\begin{table}[h!]
\begin{center}
\begin{tabular}{|c|ccccc|cc||cc|cc|}
\hline
Fields & $L$ & $E^c$ & $Q$ & $U^c$ & $D^c$ & $H^u$ & $H^d$ & $T$ & $T^c$ &
$\sigma$ & $\sigma'$\\
\hline
$SU(2)_L$ & 2 &1 &2 &1 &1 &2 &2  & 1 &1 &1 &1 \\
$A_4$ &3 &3 &3 &3 &3 &3 &3 &1 &1&3 &3\\
$Z_4$ &1 &1 &1 &1 &1 &1 &1 &$\omega$ &$\omega^3$ & $\omega^3$ & $\omega$\\
\hline
\end{tabular} \end{center}
\caption{Field content and quantum numbers of the model.}\label{tabf}
\end{table}

The most general Yukawa superpotential allowed by the above symmetry is
\begin{eqnarray}
w&=&w_Y+w_T,\\ \label{yukawa1}
{ w}_{Y}&=&y^u_{ijk} Q_i H^u_j U^c_k +
y^d_{ijk} Q_i H^d_j D^c_k +
y^l_{ijk} L_i H^d_j E^c_k,\\
{w}_{T}&=& M T T^c+ X T U^c_i \sigma_i+
\frac{Y}{\Lambda} Q_i
(H^u \cdot \sigma')_i T^c,\label{yukawanew}  
\end{eqnarray}
where $X,~Y$ and $y^{u,d,l}_{ijk}$ are the Yukawa couplings and we
assume all $y^{u,d,l}_{ijk}$ to be real for simplicity. The product of
two $A_4$ triplets denoted as $(\phi\cdot\chi)$ also transforms as a
triplet. In the above Yukawa superpotential, we consider a dimension-5
non-renormalizable operator allowed by the symmetry of model and
required for realistic quark masses and mixings as discussed later in
this section.  Such operator can arise from a renormalizable
superpotential by introducing extra messenger fields such as, for
instance, $h^u$ and $h^d$ with the same quantum numbers of $H^{u,d}$
under $SU(2)_L\times U(1)$ and transforming respectively as $\omega$
and $\omega^3$ under $Z_4$. Then the extra terms $Qh^uT^c$,
$H^uh^d\sigma'$ are allowed and, of course, the mass term
$h^uh^d$. The dimension-5 operator in $w_T$ arises after integrating
out the heavy messenger fields $h^u$ and $h^d$ from the spectrum. 
The $Z_4$ symmetry forbids all the other dimension-5 operators in the
above superpotential. In addition, we neglect corrections coming from
operators of dimension-6 or greater. An alternative could be to replace $(H^u \cdot \sigma')$
with $H^{u'}$ transforming as $\sigma'$ under $Z_4$ but as a Higgs doublet under SU(2). 
For anomaly cancellation we must also add an $H^{d'}$ with opposite charges with respect to $H^{u'}$.
One may forbid the higher order operators by replacing $Z_4$ with 
$Z_N$ symmetry with sufficiently
large $N$. We first review the main phenomenological consequences
of the $w_Y$ part of the superpotential $w$ and in the next section we
consider the $w_T$ piece, which is the new part of the present
work. By using $A_4$ product rules in the superpotential $w_Y$ it is
straightforward to show that the charged fermion mass matrices take
the following universal structure~\cite{Morisi:2009sc} 
\begin{equation}
M_{f}=\left(
\begin{array}{ccc}
0 & y_1^f \vev{ H_3^f } & y_2^f \vev{ H_2^f } \\
y_2^f \vev{ H_3^f } & 0 & y_1^f \vev{ H_1^f } \\
y_1^f \vev{ H_2^f } & y_2^f \vev{  H_1^f } & 0
\end{array}
\right),
\label{eq:me}
\end{equation}
where $f=l,u,d$ denotes charged leptons, up-type or down-type quarks
respectively and $y_{1,2}$ are the only $A_4$ invariant contractions
of the $y_{ijk}^f$. Following \cite{Morisi:2011pt} we assume the
general alignment form
\begin{eqnarray}
\label{eq:minima}
\vev{H^u}=(v^u,\varepsilon_1^u,\varepsilon_2^u)^T~~{\rm and}~~
\vev{H^d}=(v^d,\varepsilon_1^d,\varepsilon_2^d)^T~,
\end{eqnarray}
where $\varepsilon_{1,2}^u\ll v^u$ and $\varepsilon_{1,2}^d\ll v^d$,
namely $A_4$ is completely broken as a result of explicit $A_4$
soft--breaking terms in the scalar potential (see for example, the
discussion below Eqs. (5) in \cite{Morisi:2011pt}). Note that, in
addition to the ``texture'' zeros in the diagonal, one has additional
relations among the off-diagonal elements in $M_{f}$.  This may be
seen explicitly by rewriting Eq.~(\ref{eq:me}) as
\begin{equation}
M_{f}= \left(
\begin{array}{ccc}
0 & a^f \alpha^f  & b^f  \\
b^f\alpha^f  & 0 & a^f r^f \\
a^f  & b^f r^f & 0
\end{array}
\right),
\label{Mf}
\end{equation}
where $a^f=y_1^f\varepsilon_1^f$, $b^f=y_2^f\varepsilon_1^f$, with
$y_{1,2}^f$ denoting the only two couplings arising from the
$A_4$-tensor in Eq.~(\ref{yukawa1}), $r^f=v^f/\varepsilon_1^f$ and
$\alpha^f=\varepsilon_2^f/\varepsilon_1^f$.  Thanks to the fact that
the same MSSM Higgs doublet $H^d$ couples to the lepton and to the
down-type quarks one has, in addition, the following relations
\begin{equation}\label{rel}
r^l=r^d,\qquad \alpha^l=\alpha^d,
\end{equation}
involving down-type quarks and charged leptons.

Each of the mass matrices in Eq.\,(\ref{Mf}) depends on just four parameters. We can express three of the parameters
like for instance $r^f,a^f,b^f$ in terms of the corresponding fermion masses and $\alpha^f$. The results are
\begin{eqnarray}
r^f&\approx& \frac{m^f_{3}}{\sqrt{m^f_{1}m^f_{2}}}\sqrt{\alpha^f} \label{ralpha}\,,\\
a^f&\approx& \frac{m^f_{2}}{m^f_{3}}\sqrt{\frac{m^f_{1}m^f_{2}}{\alpha^f}},\label{af}\,\\
b^f&\approx& \sqrt{\frac{m^f_{1}m^f_{2}}{\alpha^f}} \label{bf}~.
\end{eqnarray}
where we have used the approximation $r^f\gg 1$ and $r^f\gg b^f/a^f$. From Eqs.\,(\ref{rel}) and (\ref{ralpha}) we
obtain the mass relation given in Eq.\,(\ref{eq:ours}). The diagonalization of $M_{d,l}$ and its phenomenological
implications have already been discussed in Ref.~\cite{Morisi:2011pt}. For example, one finds
\begin{eqnarray}
V^f_{12}&\approx&\sqrt{\frac{m_1}{m_2}}\frac{1}{\sqrt{\alpha^f}}, \label{V12} \\
V^f_{13}&\approx& \frac{m_2}{m_3^2}\sqrt{m_1 m_2}\frac{1}{\sqrt{\alpha^f}}, \label{V13} \\
V^f_{23}&\approx& \frac{m_1 m_2}{m_3^2} \frac{1}{\alpha^f}.\label{V23}
\end{eqnarray}
Using the above relations and the experimental values of the fermion masses we have $V^f_{12}
\sim \mathcal{O}(\lambda_C)$, while $V^f_{13}$ and $V^f_{23}$ are too small. In order to generate
adequate predictions for all entries of the quark mixing matrix we must modify the above scheme, as discussed in the
next section.  Note that we modify only the up-quark sector in order to maintain our mass relations
in Eq.\,(\ref{eq:ours}).

\subsection{Quark mixing}
\label{sec:quark-mixing}

In the previous section we have only considered the $w_Y$ superpotential. We now take into account also the $w_T$
superpotential terms. Assuming that the flavon fields $\sigma$ and $\sigma'$ take complex vacuum expectation values
in completely random directions, we have
\be \label{sigmavevs}
\vev{\sigma}=(\vev{\sigma_1},\vev{\sigma_2},\vev{\sigma_3})^T
\quad{\rm and}\quad
\vev{\sigma'}=(\vev{\sigma'_1},\vev{\sigma'_2},\vev{\sigma'_3})^T.
\ee
The resulting $4\times 4$ up-type quark mass matrix is 
\begin{equation}
\label{Mu}
M_{u}= \left( \ba{cccc} 0& a^u \alpha^u & b^u & Y_1 \\
  b^u \alpha^u & 0&  a^u r^u & Y_2 \\
  a^u & b^u r^u &0 & Y_3 \\
  X_1 & X_2 & X_3 & M \\ \ea \right) 
\end{equation}
where $X_i=X \vev{\sigma_i}
$ and $Y_i= Y \vev{(H^u\cdot \sigma')_i}/\Lambda$ are complex
parameters. The down quark mass matrix is unchanged and is given by
Eq.\,(\ref{Mf}).  Here we consider in detail the role of the $(T,T^c)$
coupling in the up quark sector and how it makes it possible to
account for the full structure of the CKM mixing matrix. This
possibility has already been studied in the literature in the general
case of flavour-blind models~\cite{branco:1986my,delAguila:1985ne}.
Here we explore its role in the context of our flavor--symmetric
model.
In Sec.~\ref{sec:phen-quark-sect}, we give an example set of the above
parameters which produces a viable quark mixing pattern without
spoiling the charged leptons and down-type quarks mass relations.
The idea is that the $3\times 3$ sub-matrix of the $4\times 4$ matrix
which diagonalizes on the left the up--quark mass matrix, is not
unitary.  Unitarity deviations modify the relations (\ref{V12}),
(\ref{V13}) and (\ref{V23}) allowing for an acceptable fit of the CKM
matrix. In the next section we discuss this in details.

\subsection{Neutrino masses and mixing}
\label{sec:neutr-mass-mixing}

\begin{figure}[ht!] 
 \centering
 \includegraphics[width=0.43\textwidth]{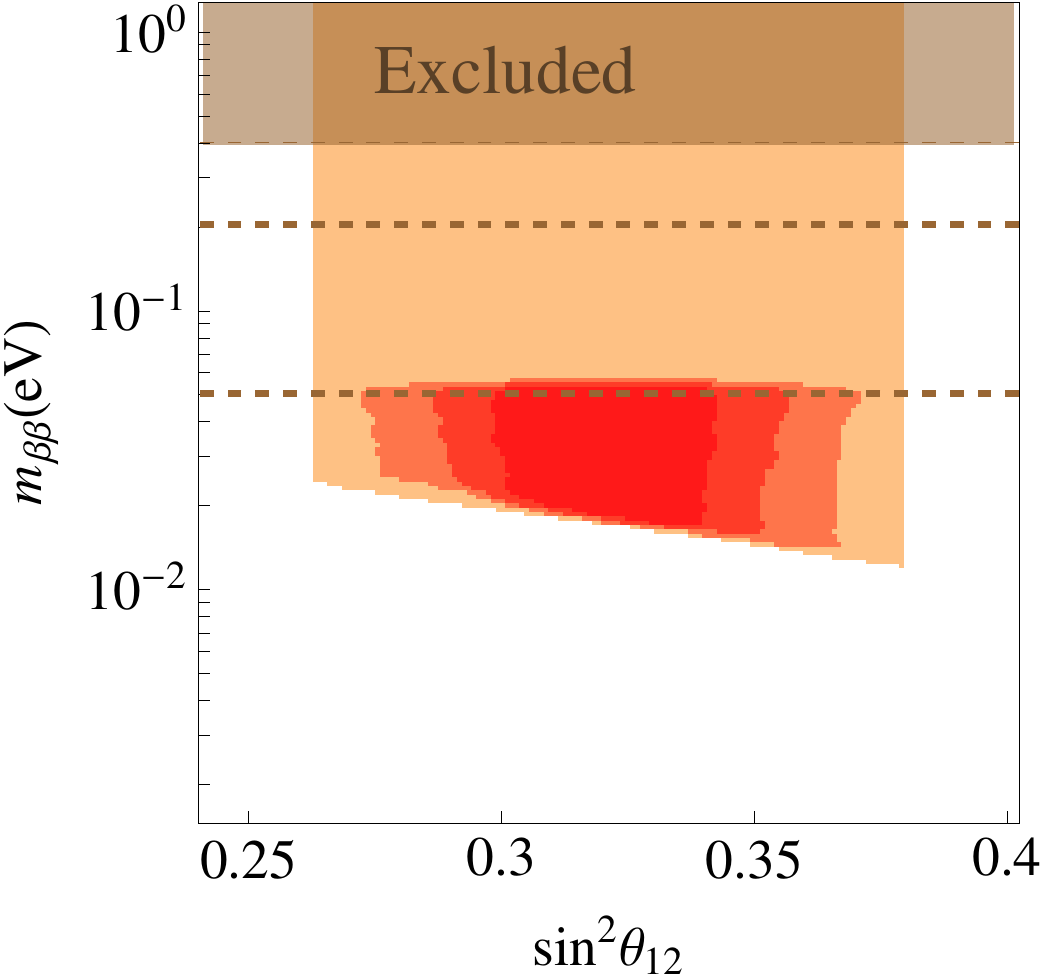}
 \includegraphics[width=0.43\textwidth]{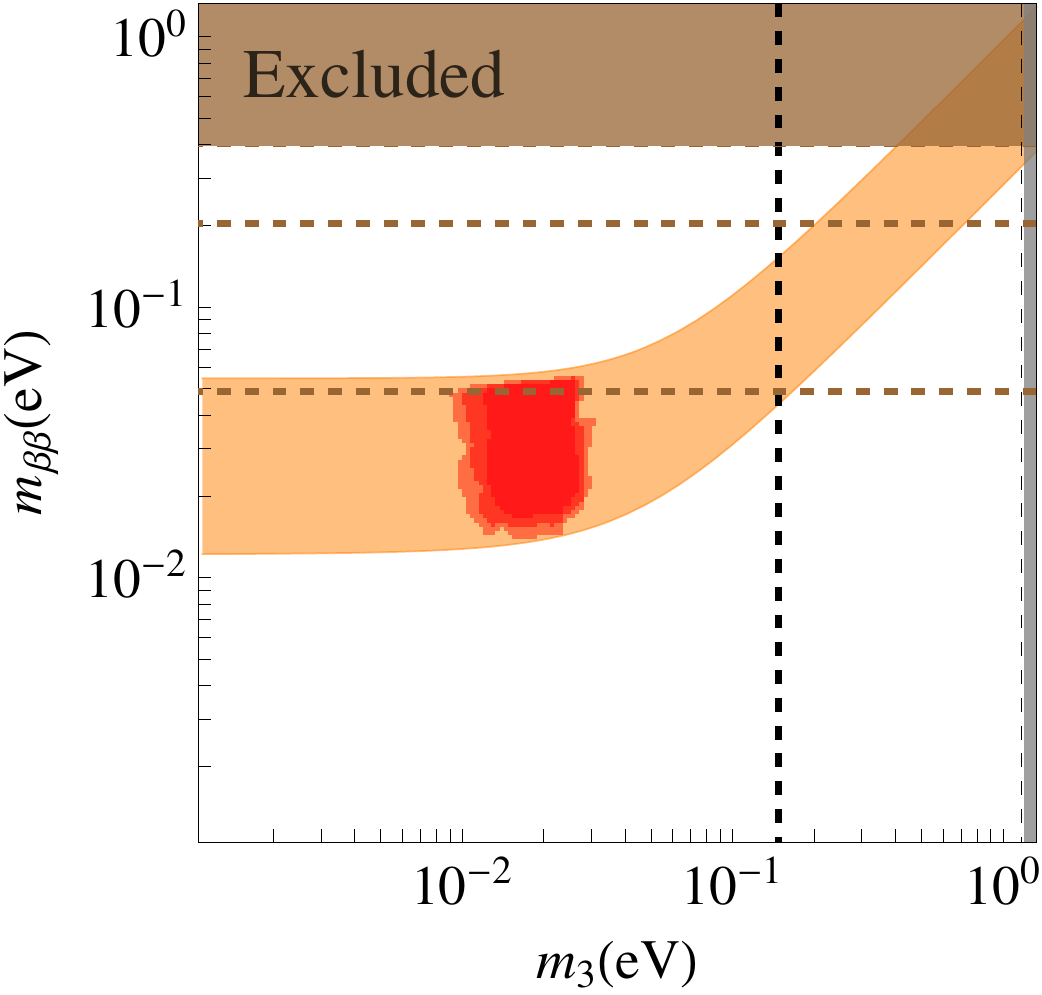}
\caption{Correlations between the neutrinoless double beta decay effective mass parameter $m_{\beta\beta}$ and the solar
mixing angle (left panel) and the lightest neutrino mass (right panel). The red regions correspond to the model
prediction while the orange ones correspond to the generically allowed regions for an inverted mass hierarchy. Darker to
lighter tones correspond to 68\%, 95\% and 99\% CL regions allowed by the model parameters. The dark brown region
corresponds to the present experimental bound on $m_{\beta\beta}$. The upper and lower horizontal dashed lines
correspond to the sensitivities of GERDA-I and CUORE experiments while the vertical dashed line in right panel shows the
future sensitivity of KATRIN experiment.}
\label{mbb}
\end{figure}

We now turn to neutrino mass generation, describing it effectively
\textit{a la Weinberg}.  In this model neutrino masses are generated
by the following dimension-5 operator
\begin{equation}
w_\nu=\frac{1}{\Lambda}L L H_u H_u.
\end{equation}
This operator is given by the product of four $A_4$-triplets and contains many possible $A_4$ contractions.  The general
structure has already been studied in Ref.\,\cite{Morisi:2009sc} and is given by
\begin{equation}
M_\nu\approx\left(\begin{array}{ccc}
x & \kappa  & \kappa \alpha^u \\
\kappa  & y & 0  \\
\kappa \alpha^u  & 0  & z
\end{array}
\right),
\label{mnu2}
\end{equation}
where $x$, $y$, $z$ and $\kappa$ are proportional to couplings that arise from each $A_4$ contraction. Note that $M_\nu$
is invariant under the $\mu$-$\tau$ interchange symmetry if $y=z$ and $\alpha^u=1$. Following \cite{Gupta:2013it}, one
can quantify the $\mu$-$\tau$ breaking as
\begin{equation} \label{mt-breaking}
 \epsilon_1 = \left\lvert\frac{1-\alpha^u}{1+\alpha^u}\right\rvert~~{\rm and}~~ \epsilon_2 =
\left\lvert\frac{y-z}{y+z}\right\rvert \,.
\end{equation}
The large but non-maximal atmospheric mixing angle as evidenced from
the recent global fits to oscillation data \cite{Tortola:2012te}
requires $y\approx z$. Interestingly, the parameter $\alpha^u$ enters
also in the up quark mass matrix. We find from the quark sector (see
section \ref{sec:phen-quark-sect} for the details) that $\alpha^u$
lies in the range $[0.7,~1.8]$ at 3$\sigma$. This implies $\epsilon_1$
in the range [0,~0.3] which corresponds to small $\mu$-$\tau$
breaking. As it is shown in \cite{Gupta:2013it}, given the large value
of $\theta_{13}$ such a small breaking of $\mu$-$\tau$ symmetry is
only consistent with either inverted or quasi-degenerate mass spectrum
of neutrinos. Moreover, the neutrino mass matrix has one zero which
forbids quasi-degenerate neutrinos and leads to a correlation between
the effective mass parameter characterizing the \znbb decay amplitude
$m_{\beta\beta}$ and the solar mixing angle as displayed in
Figure~\ref{mbb}. For convenience, we also plot $m_{\beta\beta}$ as a
function of the lightest neutrino mass eigenvalue. Clearly, the model
predicts $m_{\beta\beta} \in [0.01,~0.03]$ eV which could be
accessible in the next generation experiments. Note that we have
performed a numerical exploration of the parameter space, in which we
have also included the correction associated to the diagonalization of
the charged lepton sector and given from
Eqs.~(\ref{V12}),\,(\ref{V13}),\,(\ref{V23}).

\section{Phenomenology of the quark sector}
\label{sec:phen-quark-sect}

We now analyze the implications of the structure of the mass matrices for the quark sector. To do so we will first
remind the modifications that arise in the couplings to $W$ and $Z$ bosons. Having set the stage, we will then give an
overview of both the most relevant phenomenological consequences that should be explored and the experimental
constraints to be considered. We will then show that the model can comply with present constraints. Since in some
interesting observables this is no obstacle to accommodate predictions different from SM expectations, we will also pay
some attention to such a possibility. As usual, rotating quark fields to the mass eigenstate basis yields a
clash among left-handed up and down rotation matrices, which gives rise to the appearance of a mixing matrix $V$ in the
charged-current couplings, 
\begin{equation}
\mathscr L_W = -\frac{g}{\sqrt 2} \bar{\mathbf
  u}_L\,\gamma^\mu\,V\,\mathbf{d}_L\,W_\mu\,+\,\text{h.c.}\ .%
\label{quark:cc:01}
\end{equation}
However, since now $\mathbf{d}=(d,s,b)$ and $\mathbf{u}=(u,c,t,T)$, {\it i.e.} there is an additional up-type state, the
mixing matrix is not anymore a $3\times 3$ unitary matrix, it is a $4\times 3$ matrix. This enlarged mixing matrix $V$
can be embedded in a $4\times 4$ unitary matrix $U$~\footnote{This is exactly analogous to the rectangular
 structure of neutrino mixing within seesaw schemes~\cite{Schechter:1980gr}.}.  Non unitarity of $V$ manifests in the
couplings to the $Z$, 
\be \mathscr L_Z = -\frac{g}{2
\cos{\theta_W}}\, \left[\bar{\mathbf{u}}_L\, \gamma^\mu\, (V
V^{\dagger})\, \mathbf{u}_L - \bar{\mathbf{d}}_L\,\gamma^\mu\,
\mathbf{d}_L -2 \sin^2\theta_W\, J_{em}^\mu\right]Z_\mu\ ,%
\label{quark:nc:01}
\ee The model naturally includes flavour changing neutral couplings controlled by the deviations from $3\times 3$
unitarity of the mixing matrix: 
\be U= \left(
\begin{matrix}
V_{ud} & V_{us} & V_{ub}\\
V_{cd} & V_{cs} & V_{cb}\\
V_{td} & V_{ts} & V_{tb}\\
V_{Td} & V_{Ts} & V_{Tb}
\end{matrix}\right|\left.
\begin{matrix}
\,U_{u4}\\ \,U_{c4}\\ \,U_{t4}\\ \,U_{T4}
\end{matrix}\right)\,,\quad 
U\,U^\dagger=U^\dagger\,U=\mathbf{1}\,,\qquad
V=
\begin{pmatrix}
V_{ud} & V_{us} & V_{ub}\\
V_{cd} & V_{cs} & V_{cb}\\
V_{td} & V_{ts} & V_{tb}\\
V_{Td} & V_{Ts} & V_{Tb}
\end{pmatrix}\,,\quad
(V V^{\dagger})_{ij}=\delta_{ij}-U_{i4}^{\phantom{\ast}}U_{j4}^\ast\,.%
\label{quark:Vmat:01}
\ee

Equations (\ref{quark:cc:01}) and (\ref{quark:nc:01}) are the
cornerstones to study the phenomenological consequences of the
model. According to them, from a departure from the standard $3\times
3$ unitary mixing, we could expect
\begin{itemize}

\item modified effective vertices that involve virtual up-type quarks
  (in our scheme the $T$ quark also runs in the loops),
\item tree-level flavour-changing $Z\bar{u}_iu_j$ vertex (analogous to the
  non-diagonal $Z\nu\nu$ vertex in seesaw
  models~\cite{Schechter:1980gr}).
\end{itemize}
These would mainly affect
\begin{itemize}
\item oscillations in neutral meson systems such as kaons, $B_d$,
  $B_s$ or $D^0$ mesons, yielding potential modifications of mass and
  width differences, and CP-violating asymmetries.
\item Decays, in particular \emph{rare} decays which are typically
  loop induced in the SM (through penguins or boxes, like the
  oscillations in the previous item). To mention a few, $B\to
  X_s\gamma$ (new contributions with virtual $T$ quarks in the loop),
  $B_s\to\mu^+\mu^-$ and $K^+\to\pi^+\nu\bar\nu$ (new contributions
  with virtual $T$ quarks and $Z$ flavour changing tree level
  couplings) or $t\to cZ$ ($Z$ flavour changing tree level couplings).
\end{itemize}
Observables that are, essentially, tree-level induced, could be
modified in this framework, since ``enlarging'' the mixing matrix
necessarily implies modifications of its entries; the modifications
are, nevertheless, typically small.

In addition to the observables associated to the phenomenology of
mesons, electroweak precision data is also sensitive to the
modifications that the additional up vector-like quark produce. We
will consider the oblique parameters $S$ and $T$ (since the $U$
parameter is typically of little importance).

For a more exhaustive description of the observables we refer the
reader to \cite{Botella:2008qm,Botella:2012ju} and references therein
for details and further technicalities associated to the numerical
aspects of the exploration of the physics reach of the
model~\footnote{A systematic study of the parameter space of the model
  is conducted using Markov Chain MonteCarlo techniques. Notice in
  addition that one important difference which deserves attention: the
  analyses in \cite{Botella:2008qm,Botella:2012ju} address models with
  up vector like quarks generically, directly in terms of the
  parameters that describe the mixing matrix, without an underlying
  flavour symmetry as we have here.}. In the following we will simply
collect the most relevant observables . Tables \ref{tab:quark1},
\ref{tab:quark2} and \ref{tab:quark3} display together the ranges (1)
allowed within the model, (2) allowed within the SM and (3)
experimentally determined (where available or appropriate), for a
selected set of observables.

\subsection{CKM from deviation of unitarity}
\label{sec:ckm-from-deviation}

Table \ref{tab:quark1} illustrates two important aspects~\footnote{In
  Tables \ref{tab:quark1} to \ref{tab:quark3}, an asterisk $*$ instead
  of the corresponding experimental measurements denotes that the
  quantity is either not directly measurable, or no relevant
  measurement exists yet. Physical bounds like $|V_{tb}|\leq 1$ (by
  definition) are displayed for reference with a thin vertical line.}:
(1) the mixing element $13$, $|V_{ub}|$, is in agreement with the
experimental constraint, (2) the mixing matrix departs from the
$3\times 3$ unitary case. It is important to underline that the model
can produce an adequate value for $|V_{ub}|$: the difficulty raised in
\cite{Morisi:2011pt}, where no $T$ quark was included, is therefore
cured. In addition, it might be larger than in the SM, and this is
central if one is interested in providing some relief to the
``tensions'' that, over the last few years, have arised among the
measurements of $|V_{ub}|$, the branching ratio Br$(B^+\to\tau^+\nu)$
(which is also displayed to further illustrate the issue) and the
time-dependent CP asymmetry in $B_d^0$--$\bar B_d^0$ decays to
$J/\Psi\,K_S$, $A_{J/\Psi K_S}$. Concerning the deviations from
$3\times 3$ unitarity that the model can accommodate, the physical
(rephasing invariant) phases $\beta$ and $\beta_s$ \cite{Branco:1999fs}, together with the
mixing element $|V_{tb}|$, are displayed to illustrate the possibility
of having significant departures from their SM expectations. This is
particularly evident for $\beta_s$, which is important to describe the
mixing in the $B^0_s$ -- $\bar B^0_s$ system.

\begin{table}[ht!]
\begin{center}
\includegraphics[width=0.85\textwidth]{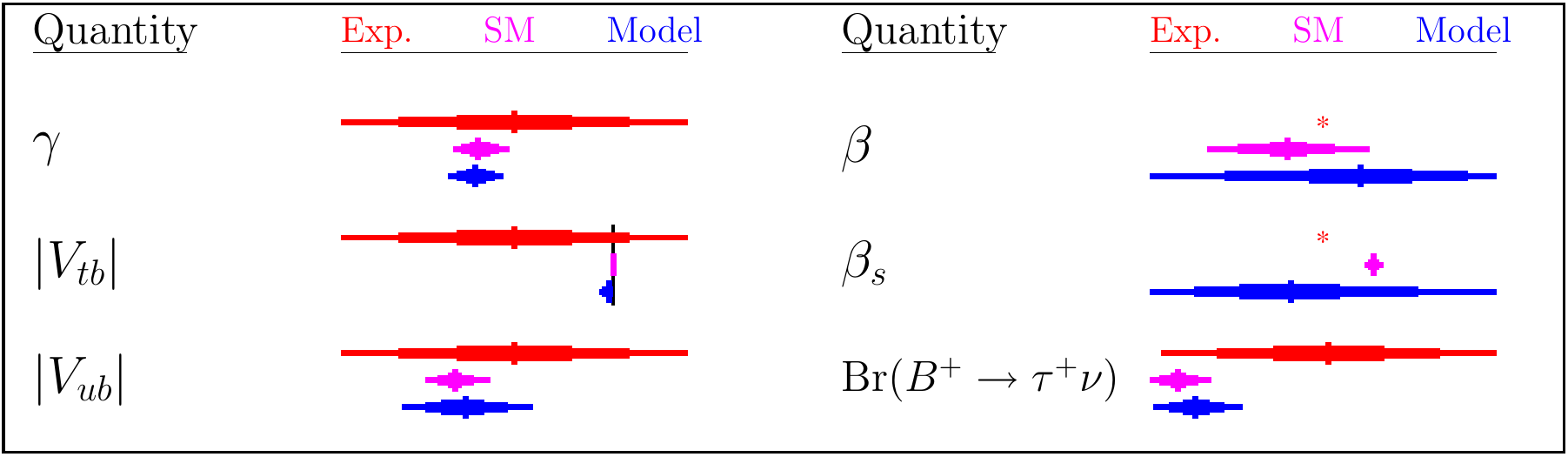}
\end{center}
\caption{Mixing elements and phases.\label{tab:quark1}}
\end{table}

\subsection{Neutral meson observables}
\label{sec:neutr-meson-observ}

Table \ref{tab:quark2} displays several observables associated to
neutral meson systems. The model agrees with the experimental
constraints. A close look to the time-dependent CP violating asymmetry
in $B^0_s\to J/\Psi \Phi$, $A_{J/\Psi \Phi}$, shows that the allowed
range in the model is much larger than the SM one: for this asymmetry
the model could accommodate values incompatible with the SM; best of
all, LHCb, which currently dominates the determination of this
asymmetry, may attain precisions sufficient to distinguish such an
hypothetical case.
\begin{table}[ht!]
\begin{center}
\includegraphics[width=0.85\textwidth]{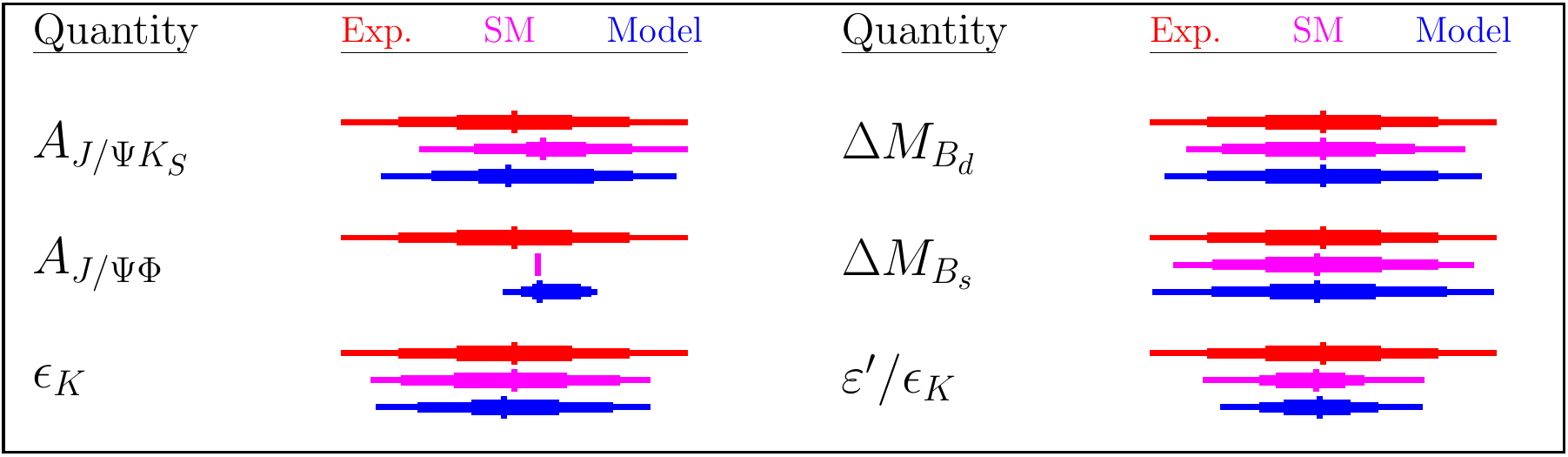}
\end{center}
\caption{Observables related to the $K^0$--$\bar K^0$, $B^0_d$--$\bar B^0_d$ and $B^0_s$--$\bar B^0_s$ systems.
\label{tab:quark2}}
\end{table}

\subsection{Rare Decays}
\label{sec:rare-decays}

While Table \ref{tab:quark2} deals with oscillation observables,
{\it i.e.} $\Delta F=2$ processes, Table \ref{tab:quark3} presents results
for some $\Delta F=1$ decays that are absent at tree level in the
SM. Accordance with the experimental constraints is again complete,
while the possibility to accommodate deviations from the SM
expectations is clearly open in several decays where experimental
progress is starting to shed light on SM territory (e.g. $K^+\to \pi^+
\nu\bar\nu$, $B_d\to\mu^+\mu^-$).
\begin{table}[ht!]
\begin{center}
\includegraphics[width=0.85\textwidth]{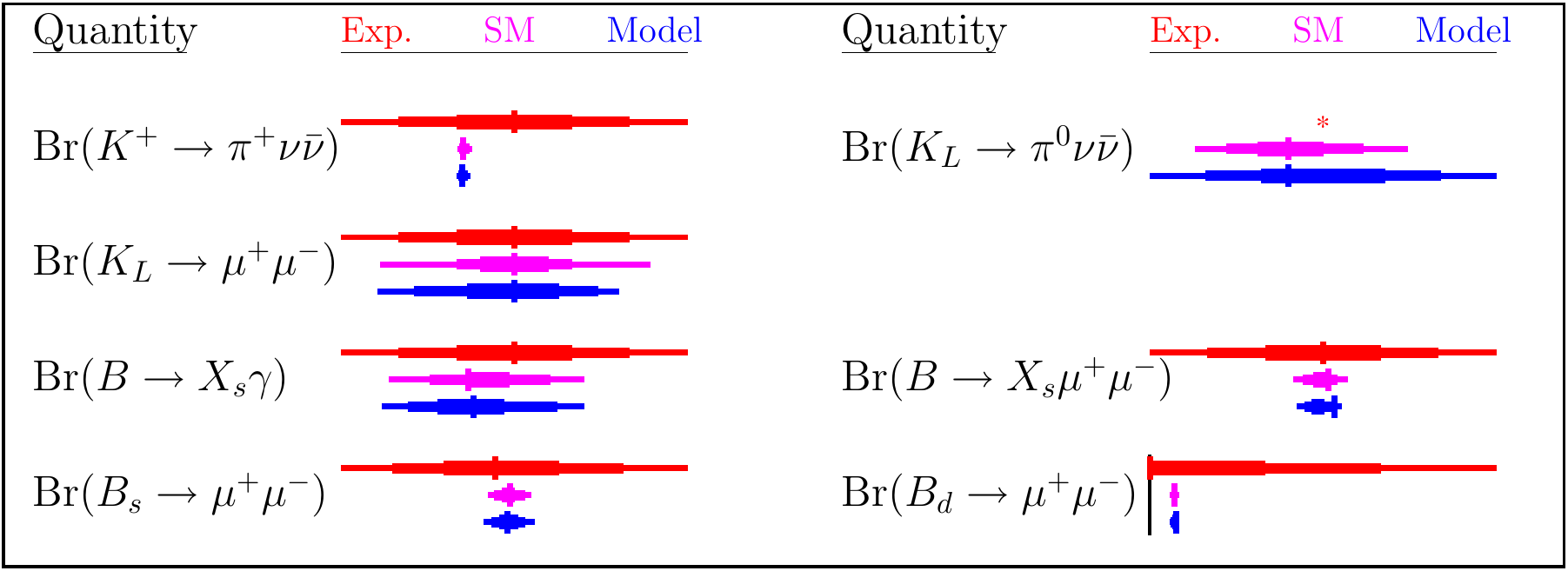}
\end{center}
\caption{Radiative decays of $K$, $B_d$ and $B_s$ mesons.\label{tab:quark3}}
\end{table}

\subsection{Additional information}

The set of observables considered above does not exhaust nor the constraints to be imposed, neither the potentially
interesting channels. Electroweak precision data has to been taken into account: to do so, agreement with the oblique
parameters $\Delta T$ and $\Delta S$ is also incorporated. Concerning beyond SM signals, rare decays
such as $t\to uZ$ and $t\to cZ$, which have highly suppressed
branching ratios in the SM, may be raised to the ${\mathcal O}(10^{-5})$
level through the Z tree level flavour changing couplings, and such
rates may be within reach of LHC experiments. Similar comments apply
to the mixing of neutral $D$ mesons. Nevertheless, since in that system
long distance hadronic interactions are relevant, we do not elaborate
and refer instead to \cite{Botella:2008qm,Botella:2012ju}. Although
further information on the model expectations can be obtained from
additional aspects like, for example, the study of correlations among
different observables or the allowed values of the mass of the new
quark $T$, Tables \ref{tab:quark1} to \ref{tab:quark3} illustrate
sufficiently that this model does indeed agree with the numerous
constraints imposed by the phenomenology of the quark sector (and may
even accommodate non standard predictions of observables to be
explored in detail in the near future, as for example the time
dependent CP asymmetry in $B_s^0\to J/\Psi\Phi$). To further confirm
this good agreement, both in the quark and the lepton sector, let us
show a specific example.

\subsection{Example}

The parameters of our example point are the following\footnote{Dimensionful parameters, that is: $|a^f|$, $|b^f|$ with
$f=l,u,d$; $|X_i|$ and $|Y_i|$ with $i=1,2,3$; $|x|$, $|y|$, $|z|$, $|k|$ and $M$, are given in GeVs.}:
\begin{align}
&r^l=r^d = 243.7417 \,,& &\alpha^l=\alpha^d =1.009666\,,& & r^u = 17226.27\,,& & \alpha^u = 1.307231\,,\nonumber\\
&|a^l| = 4.286928\cdot 10^{-4}\,, & &\arg(a^l) = 0.480176\,,& & |b^l| = 0.0072897\,, & & \arg(b^l) =
0.018797\,,\nonumber\\
&|a^u| = 5.626974\cdot 10^{-5}\,, & &\arg(a^u) = -1.486845\,,& & |b^u| = 0.009015\,, & & \arg(b^u) =
0.003155\,,\nonumber\\
&|a^d| = 2.229758\cdot 10^{-4}\,, & &\arg(a^d) = -1.248927\,,& & |b^d| = 0.012320\,, & & \arg(b^d) =0.035811\,,
\end{align}
\begin{align}
& |x|=4.792066\cdot 10^{-20}\,,&&\arg(x)=-1.697607\,,&& |y|=6.327750\cdot 10^{-20}\,, && \arg(y)=1.864509\,,\nonumber\\&
|z|=4.708915\cdot 10^{-20}\,,&&\arg(z)=1.173758\,,&& |k|=1.771198\cdot 10^{-15}\,, && \arg(k)=1.455252\,,
\end{align}

\begin{align}
&|X_1| =1.030217\,, & &\arg(X_1) =-0.215681\,,& & |Y_1| =0.549803\,, & & \arg(Y_1) =2.173908\,,\nonumber\\
&|X_2| =572.111824\,, & &\arg(X_2) =-0.756832\,,& & |Y_2| =9.278986\,, & & \arg(Y_2) =2.252815\,,\nonumber\\
&|X_3| =35.42186\,, & &\arg(X_3) =2.799661\,,& & |Y_3| =105.8904\,, & & \arg(Y_3) =-2.147215\,,\nonumber\\
&|M| =498.0447\,, & &\arg(M) =-0.368877\,.
\end{align}

The corresponding masses and mixings are:
\begin{itemize}
\item In the lepton sector:
\begin{align}
&m_e =0.5110 \text{ MeV}\,, & &m_\mu =0.1050\text{ GeV}\,,& & m_\tau =1.7768\text{ GeV}\,,\nonumber\\
&m_{\nu_1} =5.22468\cdot 10^{-2}\text{ eV}\,, & &m_{\nu_2} =5.29683\cdot 10^{-2}\text{ eV}\,,& &
m_{\nu_3} =1.60681\cdot 10^{-2}\text{ eV}\,,
\end{align}
\begin{equation}
|V_{\rm lep}|=
\begin{pmatrix}
0.80835& 0.56694& 0.15857\\
0.28458& 0.58005& 0.7632\\
0.51535& 0.58490& 0.62634
\end{pmatrix}\,.
\end{equation}

One can readily check that the mass differences are
\begin{equation}
\Delta m_{21}^2 =7.59156\cdot 10^{-5}\text{ eV}^{2}\,,\qquad \Delta m_{13}^2 =2.47155\cdot 10^{-3}\text{ eV}^{2}\,.
\end{equation}
and the mixing angles
\begin{align}
&\theta_{12}=0.611636\,, &&\theta_{23}=0.883605\,,&& \theta_{13}=0.159244\,.\nonumber\\
&\sin^2\theta_{12}=0.32971\,, &&\sin^2\theta_{23}=0.59757\,,&& \sin^2\theta_{13}=0.02514\,.
\end{align}

\item In the quark sector, with $m_T$ the mass of the new up eigenstate, 
\begin{align}
&m_d =0.002686\text{ GeV}\,, & &m_s =0.057036\text{ GeV}\,,& & m_b =3.00288\text{ GeV}\,,\nonumber\\
&m_u =0.001282\text{ GeV}\,, & &m_c =0.621607\text{ GeV}\,,& & m_t =173.1228\text{ GeV}\,,\nonumber\\
&m_T=762.93\text{ GeV}\,.
\end{align}
and
\begin{equation}
|V_{CKM}|=
\begin{pmatrix}
0.974171 & 0.225779& 0.003685\\
 0.225645& 0.973318& 0.040211\\
 0.008512& 0.040202& 0.994269\\
0.001634& 0.007730& 0.098988
\end{pmatrix}\,.
\end{equation}

\end{itemize}

\section{conclusion}
\label{sec:conclusion}

We have considered a supersymmetric model based on $A_4$ flavor
symmetry where leptons as well as quarks belong to triplets
representation of $A_4$. This kind of models predict a very
interesting mass relation between charged leptons and down quarks at
the level of the standard \SM gauge group, without unification.  In
this scenario we have studied the possibility of fitting the full
structure of quark mixing. We have introduced a pair of vector-like up
quarks.  Then the up quark mass matrix is a $4\times 4$ matrix instead
of $3\times 3$. Then the sub-block of the mixing matrix that
diagonalizes the up-quark mass violates unitary. We use such a
deviation in order to fit the $V_{ub}$ and $V_{cb}$ 
entries of the CKM matrix. A complete numerical analysis is performed to establish 
the validity of the model when relevant experimental constraints, including meson oscillations
and rare decays in kaons, $B_d$ and $B_s$ mesons, are considered. In addition, potential
deviations from SM expectations are briefly addressed.
We have also analyzed the lepton sector and found that model predicts inverted neutrino mass
spectrum leading to a potentially observable rate of neutrinoless double beta decay. 
The degree of predictivity of the model within
this sector could still be enhanced within a full-fledged seesaw-type
formulation of the model, to be taken up elsewhere.

\section*{Acknowledgments}

This work is supported by MINECO grants FPA2011-22975, FPA2011-23596 and MULTIDARK Consolider CSD2009-00064, by
Prometeo/2009/091 (Gen. Valenciana), by EU ITN UNILHC PITN-GA-2009-237920. K.M.P. acknowledges the hospitality of the
AHEP Group, IFIC, Universitat de  Val{\`e}ncia, where this work was started. S.M. supported by DFG grant WI 2639/4-1.

\appendix
\section{Experimental input}
We collect the experimental information used in the analysis in Table \ref{TAB:expdata}.
\begin{table}[!htb]
\begin{center}
\begin{tabular}{|l|c||l|c||l|c|}\hline
$\sin^2\theta_{12}$ & $0.320\pm 0.017$& $\sin^2\theta_{13}$ & $0.0248\pm 0.0029$& $\sin^2\theta_{23}$ & ($\ast$)\\
\hline
$\Delta m_{21}^2$ (eV$^2$) & $(7.62\pm 0.19)\cdot 10^{-5}$ & $|\Delta m_{13}^2|$ (eV$^2$)& $(2.50\pm
0.08)\cdot 10^{-3}$& $ m_{0\nu\beta\beta}$ (eV) & $<0.6$ eV (90\% CL)\\ \hline
$|V_{ud}|$ &$0.97425\pm 0.00022$& $|V_{us}|$ &$0.2252\pm 0.0009$& $|V_{ub}|$ & $(4.15\pm 0.49)\cdot 10^{-3}$\\ \hline
$|V_{cd}|$ &$0.230\pm 0.011$& $|V_{cs}|$ &$1.023\pm 0.036$& $|V_{cb}|$ & $0.0406\pm 0.0013$\\ \hline
$\gamma$ &$(77\pm 14)^\circ$ & $\sin 2\bar\alpha$ & $0.00\pm 0.15$& Br$(B\to\tau\nu)$ & $(11.3\pm 2.3)\cdot 10^{-5}$
\\ \hline
$\Delta M_{B_d}$ (ps$^{-1}$)& $(0.508\pm 0.004)$ & $A_{J/\Psi K_S}$ & $0.68\pm 0.02$& $\Delta \Gamma_d/\Gamma_d$ &
$-0.017\pm 0.021$\\ \hline
$\Delta M_{B_s}$ (ps$^{-1}$)& $(17.725\pm 0.049)$& $A_{J/\Psi \Phi}$ & $0.002\pm 0.087$& $\Delta\Gamma_s$ (ps$^{-1}$)
& $(0.116\pm 0.019)$\\ \hline
$A_{SL}^d$ & $-0.0030\pm 0.0078$& $A_{SL}^s$ & $-0.0024\pm 0.0063$& $A_{SL}^b$ & $-0.00787\pm 0.00196$\\ \hline
$\epsilon_K$ & $(2.228\pm 0.011)\cdot 10^{-3}$ & $\epsilon^{\prime}/\epsilon_K$& $(1.67\pm 0.16)10^{-3}$ &
\multicolumn{2}{}{} \\ \hline
{\small Br$(B\to X_s\gamma)$} & $(3.56\pm 0.25)\cdot 10^{-4}$& Br$(B\to X_s\mu^+\mu^-)$ & $(1.60\pm 0.51)\cdot
10^{-6}$& {\small 
Br$(B_s\to\mu^+\mu^-)$} & $(3.2\begin{smallmatrix}+1.5\\ -1.2\end{smallmatrix})\cdot 10^{-9}$\\ \hline
{\small Br$(K_L\to\mu^+\mu^-)$} & $(6.84\pm 0.11)\cdot 10^{-9}$& {\small Br$(K^+\to\pi^+\nu\bar\nu)$ }& $(1.73
\begin{smallmatrix}+1.15\\ -1.05\end{smallmatrix})\cdot 10^{-10}$& {\small Br$(B_d\to\mu^+\mu^-)$} & $<9.4\cdot
10^{-10}$ (95\% CL)\\ \hline
$\Delta T$ & $0.05\pm 0.12$& $\Delta S$ & $0.02\pm 0.11$ & $x_D$ & $<0.012$ (95\% CL)\\ \hline 
\cline{1-4}
\end{tabular}
\caption{Summary of experimental input \cite{Beringer:1900zz,
    Xing:2007fb}. Gaussian profiles are typically used to model the
  uncertainties.\\ ($\ast$) For $\sin^2\theta_{23}$ we use a ``double
  well'' $\Delta\chi^2$ profile, around $0.515$, to model the
  constraint taken from \cite{Tortola:2012te}.\label{TAB:expdata}}
\end{center}
\end{table}

\bibliographystyle{h-physrev4} 
\bibliography{merged}
\end{document}

%% file: bookmacros.tex
\newcommand{\bad}{\begin{array}{ccc}}
\DeclareMathAlphabet{\mathsc}{OT1}{cmr}{m}{sc}

\usepackage{color}

\def\be{\begin{equation}}        
\def\ee{\end{equation}} 
\def\bear{\be\begin{array}}       
\def\eear{\end{array}\ee} 
\def\bea{\begin{eqnarray}} 
\def\eea{\end{eqnarray}}

\def\beqa{\begin{eqnarray}}
\def\eeqa{\end{eqnarray}}

\def\beq{\begin{equation}}
\def\eeq{\end{equation}}
\def\ba{\begin{array}}
\def\ea{\end{array}}

\def\bold#1{\setbox0=\hbox{$#1$} 
     \kern-.025em\copy0\kern-\wd0 
     \kern.05em\copy0\kern-\wd0 
     \kern-.025em\raise.0433em\box0 }

\def\vev#1{\left\langle #1\right\rangle}

\newcommand {\ignore}[1]{}

\def\e6{$\mathrm{E(6)}$ }
\def\10{$\mathrm{SO(10)}$ }
\def\21{$\mathrm{SU(2)_L \otimes U(1)_Y}$ }
\def\31{$\mathrm{SU(3)_c \otimes U(1)_Q}$ }
\def\SM{$\mathrm{SU(3)_c \otimes SU(2)_L \otimes U(1)_Y}$ }

\def\3211{$\mathrm{SU(3) \otimes SU(2)_L \otimes U(1)_R \otimes U(1)_{B-L}}$ }
\def\321{$\mathrm{SU(3) \otimes SU(2) \otimes U(1)}$ }
\def\422{$\mathrm{SU(4) \otimes SU(2) \otimes SU(2)_R}$ }

\def \znbb {$0\nu\beta\beta$ }

\def\lsim{\mathrel{\rlap{\lower4pt\hbox{\hskip1pt$\sim$}}
    \raise1pt\hbox{$<$}}}         
\def\gsim{\mathrel{\rlap{\lower4pt\hbox{\hskip1pt$\sim$}}
    \raise1pt\hbox{$>$}}}         

\makeatletter
\renewcommand{\fnum@table}{\textbf{\tablename~\thetable}}
\renewcommand{\fnum@figure}{\textbf{\figurename~\thefigure}}
\makeatother

\def\mxth{\mathsurround=0pt }
\def\xversim#1#2{\lower2.pt\vbox{\baselineskip0pt \lineskip-.5pt
  \ialign{$\mxth#1\hfil##\hfil$\crcr#2\crcr\sim\crcr}}}

\def\be{\begin{equation}}
\def\ee{\end{equation}}
\def\bea{\begin{eqnarray}}
\def\eea{\end{eqnarray}}

















%% file: draft_A4-V4.bbl
\begin{thebibliography}{10}

\bibitem{Beringer:1900zz}
Particle Data Group, J.~Beringer {\em et~al.},
\newblock Phys.Rev. {\bf D86}, 010001 (2012).

\bibitem{Tortola:2012te}
D.~Forero, M.~Tortola and J.~W.~F. Valle,
\newblock Phys.Rev. {\bf D86}, 073012 (2012).

\bibitem{Morisi:2012fg}
S.~Morisi and J.~Valle,
\newblock Fortsch.Phys. {\bf 61}, 466 (2013), [1206.6678].

\bibitem{bandyopadhyay:2007kx}
ISS Physics Working Group, A.~Bandyopadhyay {\em et~al.},
\newblock Rept.Prog.Phys. {\bf 72}, 106201 (2009), [0710.4947].

\bibitem{nunokawa:2007qh}
H.~Nunokawa, S.~J. Parke and J.~W.~F. Valle,
\newblock Prog. Part. Nucl. Phys. {\bf 60}, 338 (2008).

\bibitem{ishimori2012introduction}
H.~Ishimori {\em et~al.},
\newblock {\em An Introduction to Non-Abelian Discrete Symmetries for Particle
  Physicists}Lecture Notes in Physics (Springer, 2012).

\bibitem{Babu:2002dz}
K.~S. Babu, E.~Ma and J.~W.~F. Valle,
\newblock Phys. Lett. {\bf B552}, 207 (2003), [hep-ph/0206292].

\bibitem{Altarelli:2005yp}
G.~Altarelli and F.~Feruglio,
\newblock Nucl.Phys. {\bf B720}, 64 (2005), [hep-ph/0504165].

\bibitem{Morisi:2011pt}
S.~Morisi, E.~Peinado, Y.~Shimizu and J.~Valle,
\newblock Phys.Rev. {\bf D84}, 036003 (2011), [1104.1633].

\bibitem{Georgi:1979df}
H.~Georgi and C.~Jarlskog,
\newblock Phys. Lett. {\bf B86}, 297 (1979).

\bibitem{Gatto:1968ss}
R.~Gatto, G.~Sartori and M.~Tonin,
\newblock Phys. Lett. {\bf B28}, 128 (1968).

\bibitem{King:2013hj}
S.~King, S.~Morisi, E.~Peinado and J.~Valle,
\newblock 1301.7065.

\bibitem{Morisi:2009sc}
S.~Morisi and E.~Peinado,
\newblock Phys. Rev. {\bf D80}, 113011 (2009), [0910.4389].

\bibitem{branco:1986my}
G.~C. Branco and L.~Lavoura,
\newblock Nucl. Phys. {\bf B278}, 738 (1986).

\bibitem{delAguila:1985ne}
F.~del Aguila, M.~Chase and J.~Cortes,
\newblock Nucl.Phys. {\bf B271}, 61 (1986).

\bibitem{Gupta:2013it}
S.~Gupta, A.~S. Joshipura and K.~M. Patel,
\newblock 1301.7130.

\bibitem{Schechter:1980gr}
J.~Schechter and J.~W.~F. Valle,
\newblock Phys. Rev. {\bf D22}, 2227 (1980).

\bibitem{Botella:2008qm}
F.~J. Botella, G.~C. Branco and M.~Nebot,
\newblock Phys.Rev. {\bf D79}, 096009 (2009), [0805.3995].

\bibitem{Botella:2012ju}
F.~Botella, G.~Branco and M.~Nebot,
\newblock JHEP {\bf 1212}, 040 (2012), [1207.4440].

\bibitem{Branco:1999fs}
G.~C. Branco, L.~Lavoura and J.~P. Silva,
\newblock Int.Ser.Monogr.Phys. {\bf 103}, 1 (1999).

\bibitem{Xing:2007fb}
Z.-z. Xing, H.~Zhang and S.~Zhou,
\newblock Phys.Rev. {\bf D77}, 113016 (2008), [0712.1419].

\end{thebibliography}
